\documentclass[pra,twocolumn,longbibliography]{revtex4-1}
\usepackage{graphicx}
\usepackage{amssymb}
\usepackage{bm}
\usepackage{amsmath,amsfonts,latexsym,braket}

\usepackage{hyperref}              
\usepackage{comment}
\usepackage{todonotes}

\usepackage{mathtools}

\newcommand*{\cC}{{\mathcal C}}

\newcommand*{\cL}{{\mathcal L}}
\newcommand*{\cH}{{\mathcal H}}

\newcommand*{\cS}{{\mathcal S}}
\newcommand*{\cP}{{\mathcal P}}
\newcommand*{\cO}{{\mathcal O}}
\newcommand*{\cN}{{\mathcal N}}

\DeclareMathOperator{\tr}{tr}

\DeclareMathOperator*{\E}{\mathbb{E}}

\newcommand*{\ov}[2][2.8]{\overline{\mbox{$#2$}\raisebox{#1 mm}{}}} 

\begin{document}

\title{Noisy coupled qubits: operator spreading and the Fredrickson--Andersen model}
\author{Daniel A. Rowlands}
\author{Austen Lamacraft}
\affiliation{TCM Group, Cavendish Laboratory, University of Cambridge, J. J. Thomson Ave., Cambridge CB3 0HE, UK}
\date{\today}
\email{dar55@cam.ac.uk}

\begin{abstract}
We study noise-averaged observables for a system of exchange-coupled quantum spins (qubits), each subject to a stochastic drive, by establishing mappings onto stochastic models in the strong-noise limit. Averaging over noise yields Lindbladian equations of motion; when these are subjected to a strong-noise perturbative treatment, classical master equations are found to emerge. The dynamics of noise averages of operators displays diffusive behaviour or exponential relaxation, depending on whether the drive conserves one of the spin components or not. In the latter case, the \emph{second moment} of operators -- from which the average subsystem purity and out-of-time-order correlation functions can be extracted -- is described by the Fredrickson--Andersen model, originally introduced as a model of cooperative relaxation near the glass transition. It is known that fluctuations of a ballistically propagating front in the model are asymptotically Gaussian in one dimension. We extend this by conjecturing, with strong numerical evidence, that in two dimensions the long-time fluctuations are in the Kardar--Parisi--Zhang universality class, complementing a similar observation in random unitary circuits.
\end{abstract}

\maketitle

\section{Introduction}
 The success of microscopic models of matter hinges upon the assumption that the resulting macroscopic description is relatively insensitive to the precise disposition of the constituent particles. When applied to dynamical phenomena -- collective motion -- this assumption seems at odds with our usual understanding of generic dynamical systems: that they display chaos and an exponential sensitivity to initial conditions.

That we can derive the (deterministic) laws of hydrodynamics from the motion of gas particles and the assumption of molecular chaos shows that this contradiction is not as severe as it may at first seem. By focussing on coarse-grained variables like the average local velocity, the underlying chaotic motion fades into the background and serves only to give rise to the pressure, viscosity and other parameters of the effective description.

Nevertheless, if one believes that the butterfly effect is more than a figure of speech, something must have been lost along the way. By focusing on \emph{average} quantities, the growth of \emph{fluctuations} from the microscopic to the macroscopic is obscured. Long a part of statistical fluid dynamics \cite{Monin:2013aa,Lorenz:1969aa,Ruelle:1979aa}, these questions have only recently been taken up in quantum field theory \cite{Hayden:2007aa,Sekino:2008aa,Lashkari:2013aa,Shenker:2014ab,Shenker:2014aa,Roberts:2015aa,Roberts:2015ab,Maldacena:2016aa} and many-body physics \cite{Ho:2017aa, Khemani:2017aa,Nahum:2018aa,Keyserlingk:2018aa, Gopalakrishnan:2018aa, Lucas:2017aa, Roberts:2018aa}.

Traditionally, these fields have been concerned with averages $\langle\cO_j(t)\rangle$, and response functions $i\langle[\cO_j(t),\cO_k(0)]\rangle$ of Heisenberg picture observables $\cO_j(t)$. However, the act of taking expectations in these quantities obscures the possibility that in a given experiment we may observe a very different response in observable $\cO_j(t)$ to a perturbation coupled to observable $\cO_j(0)$. The variance of the response function defines the out-of-time-order correlation function (OTOC)
\begin{equation}\label{eq:OTOC}
  \cC_{jk}(t) \equiv \frac{1}{2}\langle\left[\cO_j(t),\cO_k(0)\right]^\dagger \left[\cO_j(t),\cO_k(0)\right]\rangle,
\end{equation}
that has been suggested as a diagnostic of many-body quantum chaos \cite{Larkin:1969aa,Roberts:2015ab,Roberts:2015aa,Maldacena:2016aa}. In the light of the above discussion, it is convenient to think of the OTOC in terms of a supersystem consisting of two independent copies of the system under consideration, and extract it from the operator $\cO_j(t)\otimes \cO_j(t)$. The duplicate system is sometimes known as the thermofield double \cite{Takahasi:1974aa}.


In recent years, OTOCs have been calculated in a variety of models, including the Sachdev--Ye--Kitaev (SYK) model \cite{Kitaev:2015aa,Polchinski:2016aa,Bagrets:2017aa}, the many-body localized phase of one-dimensional spin models \cite{Swingle:2017aa,He:2017aa,Chen:2016aa,Fan:2017aa,Huang:2017aa}, weakly interacting fermions \cite{Aleiner:2016aa, Patel:2017aa}, as well as chaotic single-particle systems \cite{Rozenbaum:2017aa}.

In this work we will be concerned with the dynamics of a system of coupled qubits (spin-1/2 objects), subject to classical noise described by a stochastic process $\eta_t$ \cite{Bernard:2017aa}. We will be concerned with the first two operator moments:
\begin{equation}
  \ov{\cO_j} \equiv \E_\eta\left[\cO_j\right],\qquad \ov{\cO_j\otimes \cO_j} \equiv \E_\eta\left[\cO_j\otimes \cO_j\right].
\end{equation}
where $\cO_j\in \{X_j,Y_j,Z_j\}$ is one of the Pauli matrices describing qubit $j$. The motivations for this study are:

\begin{enumerate}
  \item Conventional wisdom suggests that noise is antithetical to quantum coherence. On the other hand, the evolution of a quantum system in the presence of classical noise is still unitary. We will see that the expected loss of coherence is only true on average: the first and second moments have completely different behaviour.

  \item The limit of strong noise provides a controlled approximation in which we can obtain a tractable dynamics of the moments.

  \item In the era of real noisy intermediate-scale quantum computers \cite{Preskill:2018aa}, there is need to understand the dynamics of quantum information in the presence of strong noise.

\end{enumerate}

The stochastic models we introduce may be regarded as continuous-time analogues of the random unitary circuit model studied in many recent works \cite{Ho:2017aa,Mezei:2017aa,Nahum:2018aa,Keyserlingk:2018aa,Khemani:2017aa,Rakovszky:2017aa,Chan:2017aa,Jonay:2018aa,Chan:2018aa,Zhou:2018ab,Bentsen:2018aa,Sunderhauf:2018aa}. Though they share some phenomenology, our analysis of the stochastic models is completely different, being based on Lindblad equations. Expectations over stochastic trajectories are taken at the first step, and the analysis of the strong-noise limit is based on conventional many body perturbation theory. This allows any (deterministic) coupling between qubits to be taken into account -- though we focus on a Heisenberg coupling for simplicity of presentation -- and allows models with conservation of one of the spin components to be handled on the same footing.

\begin{figure}
\includegraphics[width=\columnwidth]{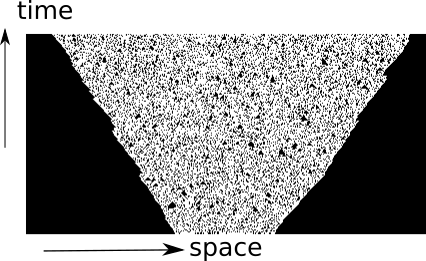}
  \caption{Growth of fronts in the one dimensional FA model with a finite average `butterfly' velocity $v_\text{B}$.}
  \label{fig:space-time-front}
\end{figure}

As is the case for the random unitary circuit models, the dynamics of moments in certain cases can be identified with the probability distribution of a classical stochastic process \cite{Oliveira:2007aa,Dahlsten:2007aa,Znidaric:2007aa,Znidaric:2008aa} (see Table \ref{table:results}). For the second moment (or OTOC) in a model without any conserved quantities, this is the Fredrickson--Andersen (FA) model, originally introduced to describe dynamics at the glass transition \cite{Fredrickson:1984aa}. The FA model is an example of a \emph{kinetically constrained model} -- see \cite{Garrahan:2011aa} for a recent review -- and has a rich phenomenology that we apply to the stochastic spin model \footnote{We note that \emph{deterministic} kinetically constrained models recently appeared in the study of certain unitary circuits \cite{Gopalakrishnan:2018aa}
}. Specifically, we will see that with appropriate initial conditions the FA model describes ballistically growing fronts that are associated with the spreading of the support of local operators in the Heisenberg picture (see Fig.\ref{fig:space-time-front}). A characteristic speed for operator spreading in many-body systems was first identified in \cite{Shenker:2014ab}, and it has since become known as the `butterfly velocity' $v_\text{B}$.

\subsection{Outline}

The outline of this work is as follows. Section \ref{sec:models} describes the models we use, and presents our derivations of the mappings to stochastic models in the strong-noise limit. In Section \ref{sec:fronts} we study the behaviour of fronts in the FA model in one and two dimensions. In Section \ref{sec:applied} we then apply the resulting phenomenology to OTOCs and the behaviour of subsystem purity. Our conclusions are discussed in Section \ref{sec:conclusions}. Technical details related to numerical simulations and calculation of the effective Lindbladians may be found in the appendices.

\section{Models and mapping}\label{sec:models}

\subsection{Models}

We consider a system of $N$ spin-1/2 objects (qubits), and the closely related cases of (1) Heisenberg picture evolution of observables $\cO(t)$: $\partial_t \cO(t) = i[H(\eta_t),\cO(t)]$ under a Hamiltonian $H(\eta_t)$ depending on a stochastic process $\eta_t$, and (2) the von Neumann equation for the density matrix $\rho(t)$: $\partial_t \rho(t) = -i[H(\eta_t),\rho(t)]$.

We will be concerned with the first two operator moments:
\begin{equation}
  \ov{\cO} \equiv \E_\eta\left[\cO\right],\qquad \ov{\cO\otimes \cO} \equiv \E_\eta\left[\cO\otimes \cO\right].
\end{equation}
The second (and higher) moments of an operator carry a great deal more information about the dynamics than the average alone. In particular, the second moment gives us:

\begin{enumerate}
  \item The average purity
  \begin{equation}
    \ov[2]{\gamma} \equiv \ov[3.3]{\tr\left[\rho_A^2\right]},
  \end{equation}
   where $\rho_A$ is the reduced density matrix of a subsystem $A$. Labelling bases of the subsystem $A$ and its complement $A^c$ by the (multi-)indices $\mathsf{A}$ and $\mathsf{A}^c$, we have
   \begin{equation}
    \ov[2]{\gamma} = \sum_{\mathsf{A}^{\phantom{c}}_1,\mathsf{A}^{\phantom{c}}_2,\mathsf{A}^c_1,\mathsf{A}^c_2} \ov[2]{\rho_{\mathsf{A}^{\phantom{c}}_1\mathsf{A}^c_1,\mathsf{A}^{\phantom{c}}_2\mathsf{A}^c_1}\rho_{\mathsf{A}^{\phantom{c}}_2\mathsf{A}^c_2,\mathsf{A}^{\phantom{c}}_1\mathsf{A}^c_2}} \; ,
   \end{equation}
  which may be extracted from $\ov[2.3]{\rho\otimes\rho}$.

   \item The average OTOC
   \begin{equation}\label{eq:OTOC}
     \ov{\cC}(j-k,t) \equiv \frac{1}{2}\ov[4.5]{\tr\left[\rho\left[\cO_j(t),\cO_k(0)\right]^\dagger \left[\cO_j(t),\cO_k(0)\right]\right]},
   \end{equation}
   where $\cO_{i,j}$ are operators on qubit $j$ and $k$. $\ov{\cC}(x,t)$ may be extracted from $\ov{\cO_j(t)\otimes \cO_j(t)}$ by contracting indices with $\rho$ and $\cO_k(0)$.

\end{enumerate}

The generalization of our approach to higher moments is straightforward, and we will return to this point in the conclusion.

We will consider two models, one with conserved total $z$-component of spin (C) and one without (NC). In both cases, the Hamiltonian is the sum of deterministic and stochastic terms
\begin{align}
  H_{\text{C}} &= H + \sum_{j=1}^N \eta^j_t Z_j \tag{C}\\
  H_{\text{NC}} &= H + \sum_{j=1}^N \boldsymbol{\eta}^j_t\cdot \boldsymbol{\sigma}_j, \tag{NC}
\end{align}
where $\boldsymbol{\sigma}=(X,Y,Z)$ are the usual Pauli matrices and $\eta_t$ is  assumed to be delta-correlated noise,
\begin{equation}
  \E_\eta\left[\eta^j_t\eta^k_{t'}\right]=g\delta(t-t')\delta_{j,k},
\end{equation}
while $\boldsymbol{\eta}_t=(\eta_t^x,\eta_t^y,\eta_t^z)$ contains three independent components (assumed identical for ease of notation). $H$ is a time-independent Hamiltonian describing local coupling between the spins. Our approach is not particularly sensitive to the precise form of $H$, but for simplicity we will begin with the Heisenberg chain
\begin{equation}
  H = J\sum_{j=1}^N \boldsymbol{\sigma}_j\cdot \boldsymbol{\sigma}_{j+1}.
\end{equation}
The generalization to other models, including those in higher dimension, will be evident.

If we formally express the white noise $\eta^j_t$ as the It\^{o} differential of a Brownian motion $B_t^j$,
\begin{equation}
\int^{t'} \eta^j_{t} \; dt = \int^{t'} dB^j_t,
\end{equation}
then one finds that the generator $dH_t^{\text{C}}$ of infinitesimal stochastic unitary evolution in model C is given by \cite{Bernard:2017aa}
\begin{equation}
 dH_t^{\text{C}} = H dt + \sqrt{g}\sum_j Z_j \, dB_t^j.
\end{equation}
A straightforward exercise in It\^{o} calculus yields the equation of motion of the density matrix \cite{Adler:2000aa}
\begin{equation}
\begin{split}
  d\rho_t &= e^{-idH_t^{\text{C}}}\rho_te^{idH_t^{\text{C}}} - \rho_t = -i\left[H,\rho\right]dt \\
&- \frac{g}{2}\sum_j\left[Z_j,\left[Z_j,\rho\right]\right] dt - i\sqrt{g}\sum_j \left[Z_j,\rho_\Psi\right]dB^j_t,
\end{split}
\end{equation}
the expectation of which is thus of Lindblad form \cite{Breuer:2002aa}
\begin{equation}\label{eq:rhobarC}
 \partial_t \bar{\rho} = \overbrace{-i\left[H,\bar{\rho}\right]}^{L_H(\bar{\rho})} -  \overbrace{\frac{g}{2}\sum_j\left[Z_j,\left[Z_j,\bar{\rho}\right]\right]}^{D(\bar{\rho})}.
\end{equation}
The corresponding equation for model NC, obtained by an analogous procedure, is
\begin{equation}
\partial_t \bar{\rho}= -i[H,\bar{\rho}] - \frac{g}{2} \sum_j \sum_{a=1}^3 [\sigma_j^a,[\sigma_j^a,\bar{\rho}]].
\end{equation}
The derivation for the first moment of an operator is obtained similarly, the only differences being $\rho \to \cO$ and $dH_t \to -dH_t$. We also note that this calculation can alternatively be done, albeit less concisely, by interpreting the stochastic process $\eta_t$ in the Stratonovich sense (see e.g. Appendix A.4 of \cite{Cai:2013aa}).

\subsection{Model C at strong noise: Symmetric Exclusion Process for $\overline{\cO}$}

A key simplification occurs in the limit of strong noise ($g$ large), where the dynamics of the moments is restricted certain \emph{slow subspaces}. In the case of $\bar\cO$, this is the kernel of the dissipator $D(\bar\cO)$. For example, in Model C the dynamics of $\ov{\cO}$ is restricted to the $2^N$-dimensional diagonal matrix elements in the basis of $Z_j$ eigenstates $\ket{z_1:z_N}$ \footnote{Here $z_1:z_N$ denotes the $N$-tuple $(z_1,\cdots,z_N)$.} with $z_j=\pm 1$. That is, only the matrix elements
\begin{equation}
  \braket{z_1:z_N|\cO|z_1:z_N}
\end{equation}
survive the averaging as they are unaffected by the dephasing noise. We now show that the evolution of the probability distribution of a spin configuration $z_1:z_N$ is described by the symmetric exclusion process (SEP) \cite{Krapivsky:2010aa} (see Table \ref{table:results}).
\begin{table}[h]
\begin{tabular}{r|c|c|}
\multicolumn{1}{r}{} & \multicolumn{1}{c}{Model NC} & \multicolumn{1}{c}{Model C} \\
\cline{2-3}
 $\ov{\cO}$ & Exponential decay & Symmetric Exclusion Process\\
\cline{2-3}
$\ov{\cO\otimes \cO}$ & Fredrickson--Andersen & `Octahedral' model\\
\cline{2-3}
\end{tabular}
\caption{The behaviour of the first and second operator moments in Models NC and C in the strong-noise limit. }
\label{table:results}
\end{table}

The dynamics on the slow subspace $\cS$ can be analyzed perturbatively in $g^{-1}$ (as is done in \cite{Cai:2013aa} and \cite{Lesanovsky:2013aa}); the effective Liouvillian to leading order takes the form $\cL_{\text{eff}} = -\cP_{\cS}L_H D^{-1} L_H \cP_{\cS}$, where $\cP_{\cS}$ is the projector onto $\cS$ and $D^{-1}$ is the inverse of the restriction of $D$ to its coimage. Explicit evaluation leads to
\begin{equation}
\cL_{\text{eff}}\bar{\rho} = -\frac{J^2}{16g} \sum_j [\sigma_j^+\sigma_{j+1}^-+\text{h.c.},[\sigma_j^+\sigma_{j+1}^-+\text{h.c.},\cP_{\cS}\bar{\rho}]].
\end{equation}
Let us regard $\bar{\rho} \in \cS$ as an element $\vec{\rho}$ of a vector space over $\mathbb{R}$ (sometimes referred to as `superspace') with basis $\{\ket{1}\bra{1},\ket{0}\bra{0}\}^{\otimes N}$. Then $\cL_{\text{eff}}$ acts as a matrix $\mathsf{L}$ on $\vec{\rho}$, giving the master equation
\begin{equation}
\partial_t \vec{\rho} = \mathsf{L} \vec{\rho},
\end{equation}
with $\mathsf{L} = \frac{J^2}{g} \sum_i \left(\vec{\sigma}_{i} \cdot \vec{\sigma}_{i+1} - \openone\right)$. The corresponding effective Hamiltonian $-\mathsf{L}$ (if we think of the master equation as an imaginary-time Schr\"{o}dinger equation) coincides with that of a Heisenberg ferromagnet. Up to a constant, $\mathsf{L}$ is thus seen to be the generator of the SEP \cite{Mendonca:2013aa}. In the one-dimensional case, we have an alternative route to this result as the model is found to be integrable by means of a mapping to an imaginary-U Hubbard model \cite{Medvedyeva:2016aa}. In the strong-noise limit, the Bethe ansatz equations reduce to those of the spin-1/2 ferromagnetic Heisenberg model, from which the quadratically dispersing Liouvillian spectrum and consequent diffusive relaxation follow (as was established earlier in \cite{Eisler:2011ab} by analytic evaluation of the single-particle Green's function).

\subsection{Model NC at strong noise: Fredrickson--Andersen model for $\ov{\cO\otimes\cO}$}

The dynamics of the first moment in model NC is trivial: the slow subspace is one-dimensional (i.e. contains only the identity) and so we find fast local relaxation rather than any hydrodynamics as in model C.

The second moment of Model NC does have interesting dynamics in the presence of strong noise, however. Because the noise in this model randomizes all components of the spins, $\ov{\cO\otimes \cO}$ lives in the tensor product of the space spanned by the rotationally invariant single-site factors
\begin{align} \label{eq:NC-singlets}
  \ket{0_j}&\equiv \frac{1}{2}\openone_j\otimes\openone_j \notag\\
  \ket{1_j} &\equiv \frac{1}{6}\left[X_j\otimes X_j + Y_j\otimes Y_j + Z_j\otimes Z_j\right].
\end{align}
Any $\ov{\cO\otimes \cO}$ of this form has the expansion
\begin{equation}
  \ov{\cO\otimes \cO} = \sum_{n_1:n_N\in \{0,1\}^N} \mathsf{C}^\cO_{n_1:n_N} \ket{n_1:n_N}.
\end{equation}
Using the properties of the Pauli matrices it is easy to show
\begin{equation}
  \ov[3.3]{\tr\left[\cO^2\right]} =  \sum_{n_1:n_N\in \{0,1\}^N} \mathsf{C}^\cO_{n_1:n_N}.
\end{equation}
Since the trace of any operator product $\tr\left[\cO_1(t)\cO_2(t)\right]$ is conserved under Heisenberg evolution, we may think of $\mathsf{C}^\cO_{n_1:n_N}$ as a probability distribution (up to overall normalization) and its evolution equation
\begin{equation}
  \partial_t\mathsf{C}^\cO = \mathsf{L} \mathsf{C}^\cO
\end{equation}
as a (classical) master equation.
In Refs.~\cite{Oliveira:2007aa,Dahlsten:2007aa,Znidaric:2007aa,Znidaric:2008aa}, a related discrete-time Markov chain was obtained for the dynamics of operator moments due to randomly chosen two-qubit unitary transformations. This Markov chain was the basis of the calculations of OTOCs and purity in the random unitary circuit model in Ref.~\cite{Nahum:2018aa}.


What stochastic process is described by $\mathsf{L}$? We will see that it is the \emph{Fredrickson--Andersen} (FA) model \cite{Fredrickson:1984aa}. The FA model is defined on a lattice with sites that may either be in state 1 or 0, with pairs of neighbouring sites $j$ and $k$ undergoing the transitions
\begin{align}\label{eq:FA-rates}
  1_j1_k \xrightleftharpoons[\Gamma_1]{\Gamma_0} 1_j0_k
\end{align}
with rates $\Gamma_{0,1}$. In the stationary state, sites are independent with probability $p_{1}=\Gamma_1/(\Gamma_1+\Gamma_0)$ to be 1. We find $\Gamma_0=\Gamma_1/3=4J^2/3g$ for model NC, i.e., 1s are three times more common than 0s:
\begin{equation}\label{eq:stationary}
  \mathsf{C}^\cO_{n_1:n_N}|_\text{stationary}=\frac{1}{4^N}\prod_j 3^{n_j}.
\end{equation}
Two further comments: (1) The dynamics of $\mathsf{C}^A$ and $\mathsf{C}^\rho$ are identical because the rates are quadratic in $J$. (2) Individual trajectories of the FA model have no meaning, as only the probability distribution $C^{\cO}$ appears in the moment $\ov{\cO\otimes \cO}$.

The derivation of the effective dynamics for $\overline{\cO\times\cO}$ follows the same pattern as for the first moment. Noise averaging the stochastic differential equation for the second moment of $\rho$ in model NC leads to the Lindblad equation for the replicated system
\begin{equation} \label{eq:NC-rhorho-Lindblad}
\begin{split}
\partial_t \ov{\rho \otimes \rho} = -i[\cH,\ov{\rho \otimes \rho}] - \frac{g}{2}\sum_{j,a} [\Sigma_j^a,[\Sigma_j^a,\ov{\rho \otimes \rho}]]
\end{split}
\end{equation}
where we have introduced the operators $\cH = H \otimes \openone + \openone \otimes H$ and $\Sigma_j^a = \sigma_j^a \otimes \openone + \openone \otimes \sigma_j^a$.

The kernel of the dissipator determines the slow subspace $\cS = \mathop{\text{span}}\left(\{\ket{0},\ket{1}\}^{\otimes N}\right)$, where the $\{\ket{0},\ket{1}\}$ states were defined in \eqref{eq:NC-singlets}. The effective Liouvillian to leading order (see Appendix \ref{sec:Leff-modelC} for the derivation for model C; the model NC result is obtained similarly) acts on elements of $\cS$ according to
\begin{equation}
\cL_{\text{eff}}(\cdots) = -\frac{1}{4g} \cP_{\cS}[\cH,[\cH,\cdots].
\end{equation}
A matrix representation for $\cL_{\text{eff}}$ can again be found (see Appendix \ref{sec:Leff-modelNC} for details) if we take $\{\ket{0}, \ket{1}\}^{\otimes N}$ as a vector space basis. This matrix is given by $\mathsf{L} = \sum_j \mathsf{L}_{j,j+1}$, where
\begin{equation}\label{eq:FA-generator}
\mathsf{L}_{j,j+1}=\frac{4J^2}{g}
\begin{pmatrix}
0 & 0 & 0 & 0 \\
0 & -1 & 0 & 1/3 \\
0 & 0 & -1 & 1/3 \\
0 & 1 & 1 & -2/3
\end{pmatrix},
\end{equation}
can be identified as the transition rate matrix of a continuous-time Markov process: the one-spin facilitated FA model with rates given in Eq.~\eqref{eq:FA-rates}.

A similar analysis for the second moment of model C does not appear to lead to a mapping to a classical stochastic model, but we nevertheless make some observations about the effective dynamics in Appendix \ref{sec:Leff-modelC}.

\section{Phenomenology of fronts}\label{sec:fronts}

\subsection{Fronts in the FA model}

The FA model was originally introduced to describe dynamics at the glass transition, and is an example of a \emph{kinetically constrained model}, see \cite{Garrahan:2011aa} for a recent review. The model has a spectral gap \cite{Cancrini:2008aa}, indicating that equilibrium fluctuations are generically exponentially decaying in time. Our main interest, however, is in the nonequilibrium dynamics of the model, in particular in initial conditions with only a few 1s, or regions devoid of 1s. In this case a nonzero density of 1s grows into the empty region with a finite front velocity, see Fig.~\ref{fig:space-time-front}.

The dynamics of a front in the FA model in one dimension was recently analyzed rigorously in Ref.~\cite{Blondel:2018aa} for a variant of the model \cite{Graham:1993aa} in which the transition rate is independent of the number of neighbours (see Ref.\cite{Blondel:2013aa} for the related case of the East model). There it was shown that if the rightmost 1 starts at site $0$, its displacement $X_t$ after time $t$ is asymptotically given by the normal distribution
\begin{equation}\label{eq:CLT}
  \frac{X_t - v_\text{B}t}{\sqrt{t}}\overset{d}{\underset{t\to\infty}{\longrightarrow}} \cN(0,s^2),
\end{equation}
for some $v_\text{B}$ and $s$. This chimes with the arguments given in Refs.~\cite{Nahum:2018aa,Keyserlingk:2018aa,Khemani:2017aa} for the random unitary circuit model that the probability distribution of $X_t$ is that of a biased random walk.

\subsection{Fronts in two dimensions}

The derivation of the FA model in the strong-noise limit holds in any dimension. Ballistic motion of the front in kinetically constrained models in higher dimensions is discussed in Refs.~\cite{Leonard:2010aa,Douglass:2013aa,Gutierrez:2016aa}. It is natural to ask how the front distribution in \eqref{eq:CLT} generalizes to higher dimensions. For the random unitary circuit model, Ref.~\cite{Nahum:2018aa} proposed -- and provided numerical evidence -- that the fluctuations of the front at long times are in the universality class of the Kardar--Parisi--Zhang (KPZ) equation \cite{Kardar:1986aa,Kardar:2007aa}. In the $1+1$-dimensional case, relevant for the growth of a front in two dimensions, this equation has the form
\begin{equation}\label{eq:KPZ}
  \partial_t h = c_0 +\nu\partial_x^2h + \frac{\lambda}{2}(\partial_x h)^2 +\zeta(x,t).
\end{equation}
Here $h(x,t)$ denotes the displacement of the front in the direction of growth, as a function of transverse coordinate $x$. The first term in \eqref{eq:KPZ} is a contribution to the ballistic growth rate; the second describes diffusive motion of the surface; the third captures a quadratic dependence of the local growth rate on the tilt of the surface; the last is a spatially uncorrelated white noise. The quadratic term is a relevant perturbation below two spatial dimensions that is responsible for novel scaling behaviour. For the one-dimensional case considered here, fluctuations of the surface have a dynamical critical exponent $z$ -- describing the relative scaling of spatial and temporal fluctuations as $t\sim x^z$ -- of $z=3/2$, and growth exponent $\beta$ -- describing the growth of interface fluctuations as $h\sim t^\beta$ -- of $\beta=1/3$.

\begin{figure}
\includegraphics[width=\columnwidth]{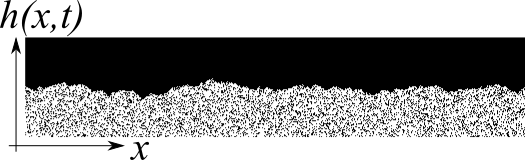}
\includegraphics[width=\columnwidth]{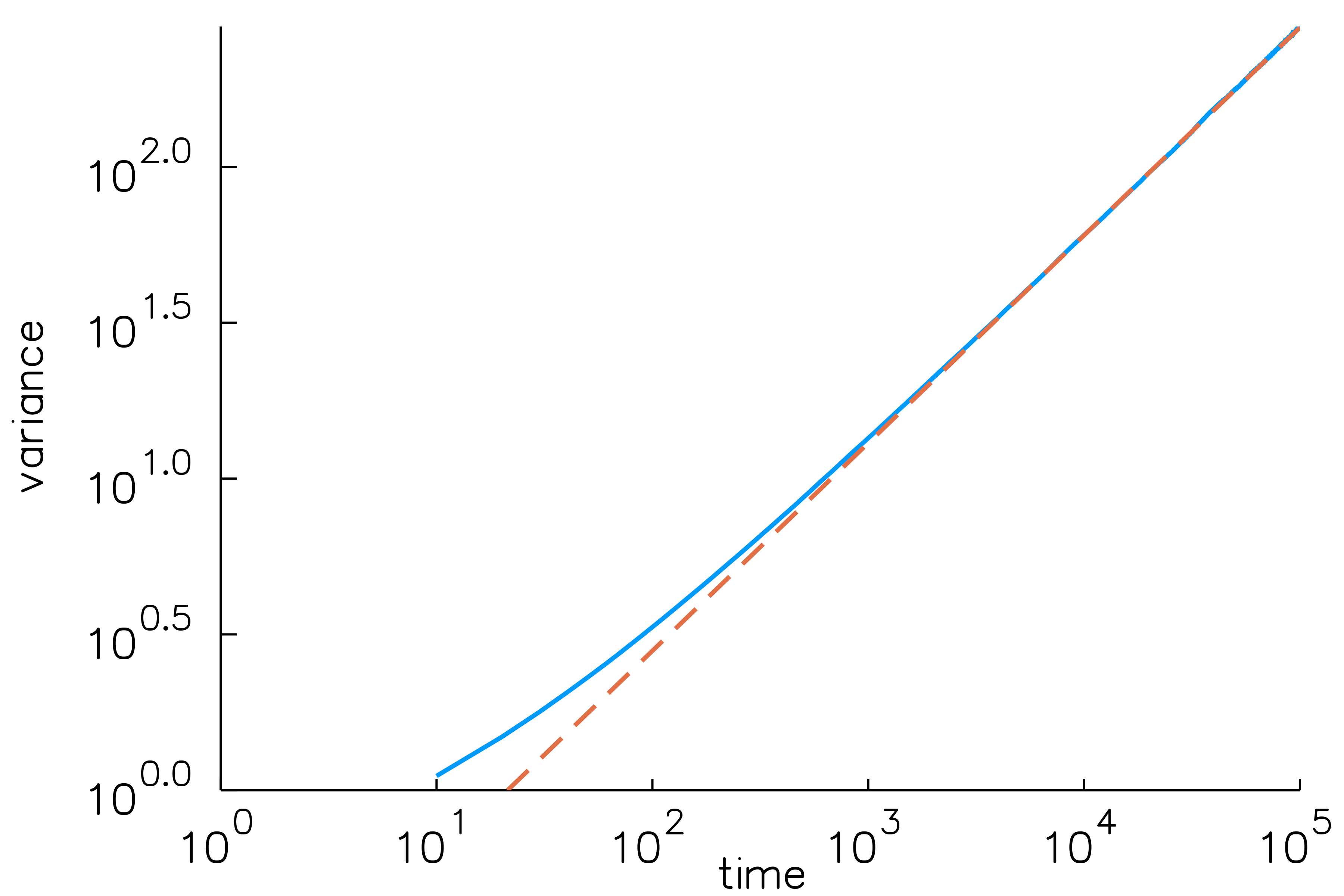}
  \caption{(Top) Upward growth of a front in the 2D FA model into a region of 0s (black), starting from a row of 1s. (Bottom) Growth of front variance with time. Dashed line is the power law $0.13 t^{2/3}$, consistent with the KPZ growth exponent $\beta=1/3$.}
  \label{fig:growth-exponent}
\end{figure}

We performed a numerical simulation to determine the growth exponent for the FA model. The details are described in Appendix \ref{sec:numerical}. For simplicity, we considered the growth from a row of 1s, corresponding to flat initial conditions, rather than from a single 1, which leads to a rounded cluster. This option was not available to the authors of Ref.~\cite{Nahum:2018aa}, as the peculiarities of the circuit model mean there is no roughening in a lattice direction. In a simulation of $10^5$ time steps, we observe nearly two decades of scaling with the KPZ exponent $\beta\sim 1/3$ (see Fig.~\ref{fig:growth-exponent}).

There is a wealth of exact results for the $1+1$-dimensional KPZ universality class: see Ref.~\cite{Kriecherbauer:2010aa} for a recent review. In particular, the long-time scaling form of the probability distributions of the height of a growing interface has been determined starting from various initial conditions. More precisely, we write
\begin{equation}
  h(x,t) \overset{d}{\underset{t\to\infty}{\longrightarrow}} ct + \alpha t^{1/3}\chi,
\end{equation}
where $\chi$ is a random variable with known distribution, and $c$ and $\alpha$ are constants. In the case of flat initial conditions, $\chi$ is drawn from the Tracy-Widom distribution corresponding to the Gaussian Orthogonal Ensemble (GOE) \cite{Prahofer:2000aa,Baik:2001aa,Tracy:1996aa}.

\begin{figure}
\includegraphics[width=\columnwidth]{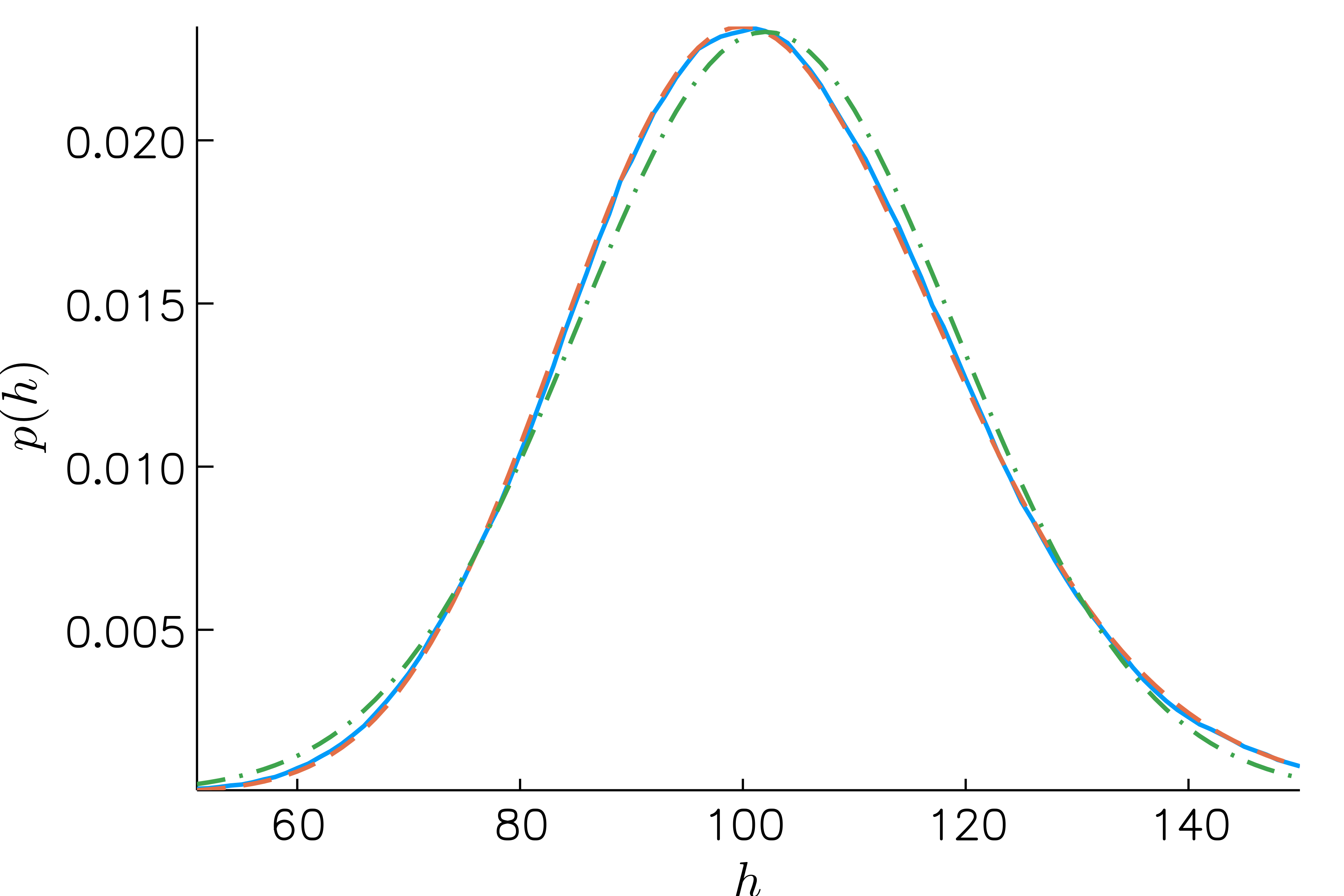}
  \caption{Front probability distribution at $t=10^5$ compared with best fit Tracy-Widom density $F_1'(s)$ (dashed) and Gaussian (dash-dotted), where the width and shift of the distributions are fitting parameters.}
  \label{fig:TW-compare}
\end{figure}

In Fig.~\ref{fig:TW-compare} we show a comparison between the probability distribution of the front obtained at the end of our numerical simulation, and the best fit Tracy-Widom GOE and Gaussian distributions. The superiority of the Tracy-Widom fit is evident, in particular in capturing the skew of the distribution and the differing behaviour of the left and right tails
\begin{equation}
  \log p(h) \propto \begin{cases}
  - |h|^3 & h\to -\infty\\
  -h^{3/2} & h\to +\infty.
\end{cases}
\end{equation}

\section{Applied Phenomenology}\label{sec:applied}

\subsection{OTOCs}

The mapping to the FA model yields a simple expression for the OTOC \eqref{eq:OTOC} in Model NC with infinite temperature initial density matrix $\rho$
\begin{equation}
\ov{\cC}_{jk}(t) = 2^{-N}\left(\frac{4}{3}\right)\sum_{n_1:n_N \in \{0,1\}^N \atop n_k=1} C_{n_1:n_N}^{\cO_j}(t),
\end{equation}
where the normalization follows from $\tr\left[\cO_j(t)^2\right]=2^N$. For a FA model starting with only site $j$ in state 1, the OTOC is then ($4/3$ times) the probability $k$ has value 1 at time $t$.

\begin{figure}
\includegraphics[width=\columnwidth]{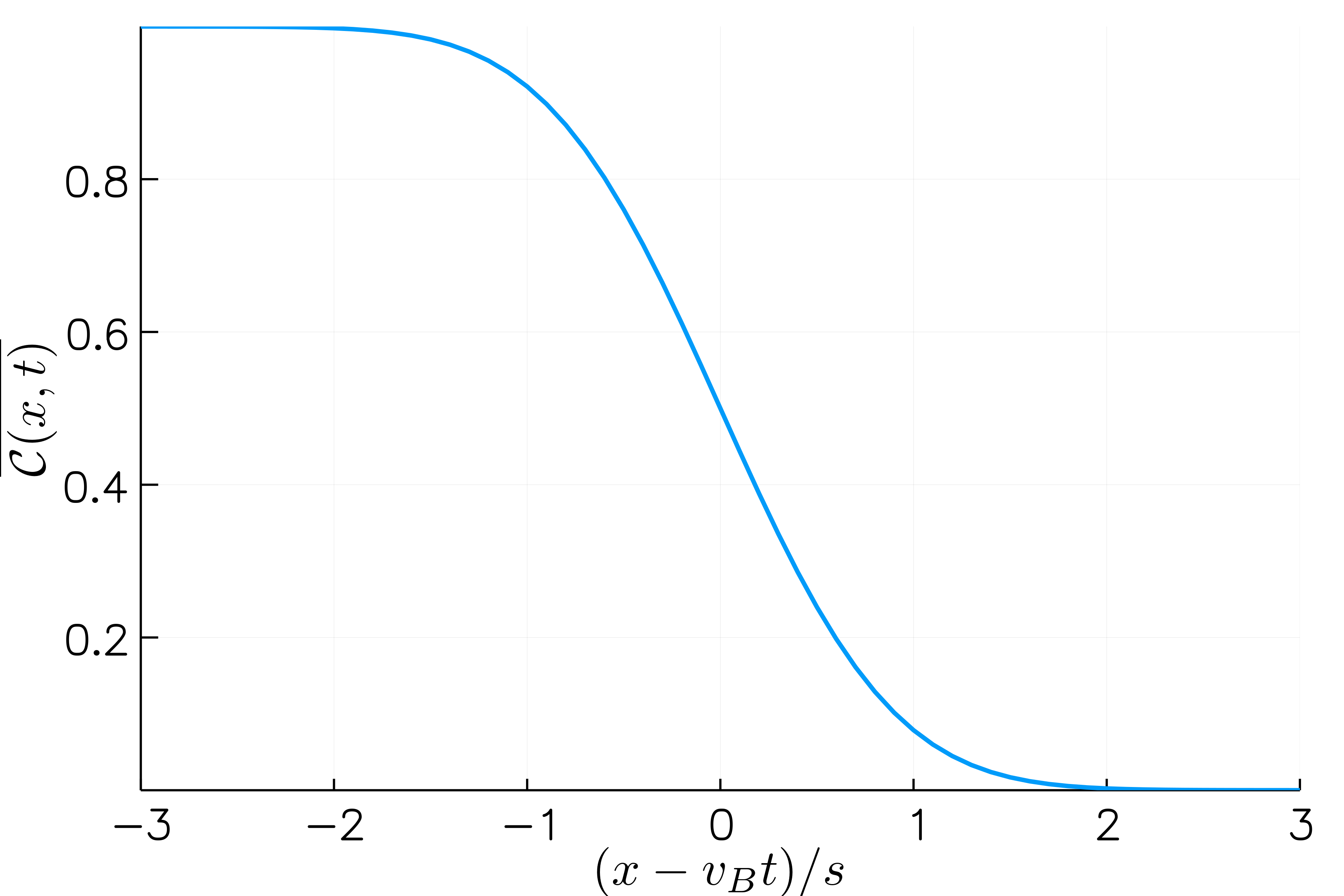}
  \caption{The averaged OTOC $\ov{\cC}_{jk}(t)$ is identified with the probability for sites $j$ and $k$ to be in the same cluster. Here, we have illustrated the Gaussian case -- the functional form being that of an error function, the cumulative distribution function of a Gaussian -- appropriate to one dimension.}
  \label{fig:OTOC}
\end{figure}

Let us further make the reasonable assumption that, after the front arrives at site $k$, the probability to be in state 1 quickly approaches the equilibrium value of $3/4$. Then we identify
\begin{equation}
  \ov{\cC}_{jk}(t) = \mathop{\text{Pr}}\left[\text{$k$ in cluster seeded by $j$}\right],
\end{equation}
or in other words, the cumulative distribution function of the front. The resulting behaviour of the OTOC is illustrated in Fig.~\ref{fig:OTOC}.

\subsection{Purity decay}

Consider a partition of the qubits into sets $A$ and $A^c$, of sizes $|A|$ and $|A^c|$. The average purity of a region $A$ is expressed as (c.f. Ref.~\cite{Mezei:2017aa,Keyserlingk:2018aa})
\begin{equation}\label{eq:purity}
  \ov[2]{\gamma}=\ov[3.3]{\tr\left[\rho_A^2\right]}=2^{|A^c|}\sum_{n_j=0,1 \text{ for } j\in A \atop n_j = 0 \text{ for } j\in A^c } \mathsf{C}^\rho_{n_1:n_N}.
\end{equation}
The purity is ($2^{|A^c|}$ times) the probability that $A^c$ contain only $0$s.
Consider taking as an initial condition a random pure product state, described by a density matrix
\begin{equation}
  \rho = \frac{1}{2^N}\prod_j \left[\openone_j + \boldsymbol{\varrho}_j\cdot\boldsymbol{\sigma}_j\right]
\end{equation}
with unit vectors $\boldsymbol{\varrho}_j$. Projected into the slow subspace of Model NC, this gives $\mathsf{C}_{n_1:n_N}(t=0)=2^{-N}$. Comparing with \eqref{eq:purity}, we see $\overline\gamma(t=0)=1$, as required. Note that the overall probability of $A^c$ being empty is $1/2^{|A^c|}$, but this exponentially small factor is cancelled by the prefactor in \eqref{eq:purity}.

\begin{figure}
\includegraphics[width=0.8\columnwidth]{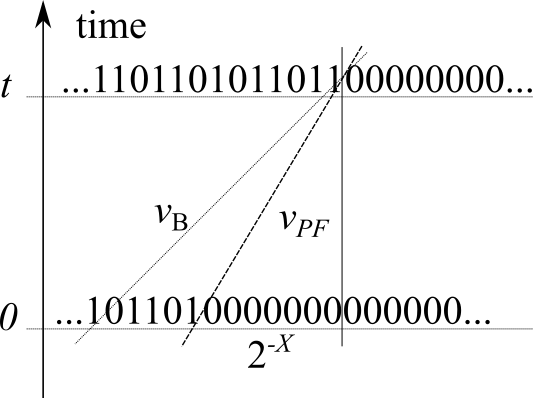}
  \caption{A `purity front' velocity $v_\text{PF}<v_\text{B}$ arises from the joint probability of the propagating front deviating from the velocity $v_\text{B}$ and an initial condition with $X$ extra 0s.}
  \label{fig:purity-velocity}
\end{figure}

If $A^c$ is empty (i.e. contains only $0$s) at time $t$, we expect the fronts to be within $A$ at earlier times. To see how this picture leads to the decay of purity, consider the growth of a single front in one dimension. The position of the front is described by \eqref{eq:CLT}. To find the most likely trajectory, the probability of finding a front at a distance $X$ inside $A$ at time 0 must be combined with the probability of the front propagating to the boundary between $A$ and $A^c$ at time $t$ (see Fig.\ref{fig:purity-velocity}). The joint probability of these two events is then
\begin{equation}\label{eq:joint}
  P(X_t=0,X_0=-X)\sim\frac{1}{2^{|A^c|+X}}\exp\left(-\frac{\left[X-v_\text{B}t\right]^2}{2s^2 t}\right).
\end{equation}
For large $t$, it suffices to find the optimum value $X_*$ of the initial front position, giving a `purity front' velocity $v_\text{PF}<v_\text{B}$
\begin{equation}
  v_\text{PF}\equiv\frac{X_*}{t} = v_\text{B} - s^2\ln 2.
\end{equation}
The fronts move slower than the butterfly velocity $v_\text{B}$. Note that a similar argument appears in Ref.~\cite{Mezei:2017aa}, though with \emph{ad hoc} assumptions about the statistics of front motion.

Substituting the optimal value $X_*$ in \eqref{eq:joint} gives the exponential decay of the purity
\begin{equation}
  \gamma(t) \sim \exp\left(-v_\text{B}t\ln{2}+\frac{s^2t}{2}\ln^2 2 \right),
\end{equation}
which enables us to define the `purity velocity' $v_{\text{P}} = v_\text{B} - \frac{1}{2}s^2\ln 2$, such that $\gamma(t) \sim \gamma(0)e^{-v_\text{P}t\ln 2}$.

\begin{figure}
\includegraphics[width=1\columnwidth]{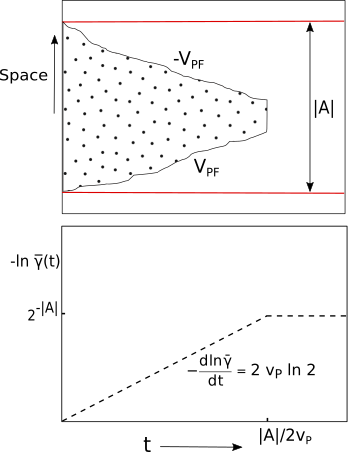}
  \caption{(Top) Schematic representation of the propagation of two fronts in 1+1 spacetime. White represents a region of 0s and black dots an active region of 1s and 0s. (Bottom) $-\ln\bar{\gamma}(t)$ for the case when $\bar{\gamma}(t)$ is computed for a finite subsystem. The red lines at fixed spatial positions in the upper right figure demarcate a finite subsystem of size $|A|$. Before saturation, $-\ln \bar{\gamma(t)}$ grows at a rate (shown) controlled by the purity velocity.}
  \label{fig:FA-front}
\end{figure}

Applied to a region $A$ of finite size, a simple generalization of the above argument implies that two fronts move towards each other at $\pm v_\text{PF}$. However, the two fronts \emph{never touch} (see Fig.~\ref{fig:FA-front}), for when $t>|A|/2v_\text{P}$, the most likely initial configuration is completely empty, and the purity saturates. Thus we have
\begin{equation}
  \gamma(t)\sim\begin{cases}
  e^{-2v_\text{P}t\ln 2} & t<|A|/2v_\text{P}\\
  \frac{1}{2^{|A|}} & t>|A|/2v_\text{P}
\end{cases}
\end{equation}
These results are valid for large $t$ and $|A|$, where the optimum dominates the probability. The fact that purity decays on a time scale linear in the subsystem size is consistent with our local Hamiltonian, and is to be contrasted with the fast scrambling (i.e. in a time logarithmic in the size) possible in systems with highly nonlocal coupling \cite{Sekino:2008aa,Lashkari:2013aa}.

If $|A^c|<|A|$, the situation is slightly different. Once $t>|A^c|/2v_\text{P}$, the most probable way for an empty $A^c$ to arise is from the stationary distribution \eqref{eq:stationary}, assuming this distribution is approached exponentially quickly from the initial state, giving
\begin{equation}
  \gamma(t>|A|/2v_\text{P})=\frac{1}{2^{|A^c|}}.
\end{equation}


The purity dynamics we have found (see Fig.~\ref{fig:FA-front}) are consistent with the expectation, largely based on exact diagonalization studies \cite{Kim:2013aa} and toy models \cite{Ho:2017aa}, that ballistic entanglement growth is a universal feature of quantum chaotic many-body systems. By Jensen's inequality, the growth rate of $-\ln \bar{\gamma}(t)$ that we have calculated is a lower bound on the growth of the averaged second R\'{e}nyi entanglement entropy. Moreover, our continuous-time results supplement the analytic discrete-time calculations in random unitary circuits \cite{Nahum:2018aa}, which have also confirmed this phenomenology (with a purity velocity satisfying $v_{\text{P}} < v_{\text{B}}$), via a mapping of the average purity onto the partition function of a directed random walk. The approach, which has been extended to obtain the higher R\'{e}nyi entropies from a correspondence with a hierarchy of classical statistical mechanics models \cite{Zhou:2018aa}, has motivated the suggestion of a ``minimal membrane'' picture of entanglement spreading in generic nonintegrable quantum systems \cite{Jonay:2018aa}.

\section{Conclusions}\label{sec:conclusions}
We have provided a precise account of operator spreading for a system of interacting qubits undergoing continuous time evolution, with each qubit independently coupled to a stochastic drive. By averaging over noise, Lindblad equations for the first and second operator moments were derived and studied perturbatively in the strong-noise limit; the central result being the identification of a mapping to the Fredrickson--Andersen model for the second moment dynamics in the case of noise that does not conserve a spin component. Considering the phenomenology of front growth in this model then enabled us to determine the implications for the behaviour of OTOCs and the decay of subsystem purity, which were found to be in line with the results established in random unitary circuit models. Although the mapping holds in arbitrary dimension, we restricted our attention to the one- and two-dimensional case: in one spatial dimension, we exploited the known exact Gaussian asymptotics of the front, whilst in two dimensions we conjectured, with numerical support, that front fluctuations exhibit (1+1)-dimensional KPZ universality, thus giving us access to exact results for the front shape in terms of Tracy-Widom distributions.

Our approach generalizes naturally to higher operator moments \cite{Cotler:2017aa}. We expect that the identification of the slow subspaces and projection of the dynamics into those spaces will be more involved but tractable, and will allow the study of higher entanglement entropies and the full distribution of operator statistics.


%

\noindent\emph{Acknowledgments}. DAR and AL gratefully acknowledge the EPSRC for financial support, under Grant Nos. EP/M506485/1 and EP/P034616/1 respectively. We thank Oriane Blondel, Juanpe Garrahan, Robert Jack and Katarzyna Macieszczak for useful discussions. This paper was finished while AL was a participant in the programme `Quantum Paths' at the Erwin Schr\"odinger International Institute for Mathematics and Physics (ESI).

\appendix

\section{Details of numerical simulation}\label{sec:numerical}

For convenience we study the version of the FA model in which the probability of a site to flip its state depends only on having neighbours, not their number \cite{Graham:1993aa}.

We use multispin coding \cite{Friedberg:1970aa,Jacobs:1981aa,Landau:2014aa}, whereby 64 configurations of the model are represented as an array of (unsigned) 64-bit integers, and updated by bitwise operations. This allows 64 trajectories to be simulated simultaneously on a single core. Our simulation consisted of a single run on each of 16 virtual cores, corresponding to 1024 trajectories.

We initialized a $L\times H$ lattice with $L$ 1s in the first row, enforcing periodic boundary conditions along the rows. The front is defined as the height of the highest $1$ in each column. As the front grows in the vertical direction, it must be periodically reset so that it remains roughly centered. This is achieved by calculating the mean height of the front across the 64 configurations every 10 updates and moving the configuration downward by 10 sites when the mean exceeds $H/2+5$.

For the largest simulations we took $L=10^4$, $H=200$, and $T=10^5$ timesteps. A height of $H=100$ resulted in a breakdown of scaling behaviour at the longest times, presumably a consequence of the fluctuations of the interface being bounded by the finite height window.

The simulation code and data analysis are written in Julia and are available as a Jupyter notebook, together with the simulation data, at \url{https://github.com/AustenLamacraft/FA-front}.

\section{Second moment dynamics of Model C in the strong-noise limit} \label{sec:Leff-modelC}
Application of It\^{o}'s lemma enables us to write down the stochastic differential equation for the second moment of the density matrix in model C
\begin{equation}
\begin{split}
d(\rho \otimes \rho) &= -i[\cH,\rho \otimes \rho]\,dt -i\sqrt{g}\sum_j [\Sigma_j^z,\rho \otimes \rho]\, dB_t^j  \\
&-\frac{g}{2} \sum_j  [\Sigma_j^z,[\Sigma_j^z,\rho \otimes \rho]] \, dt,
\end{split}
\end{equation}
which upon averaging leaves us with
\begin{equation} \label{eq:rhorho-eqofmotion}
\partial_t \ov{\rho \otimes \rho}  = -i [\cH, \ov{\rho \otimes\rho}] - \frac{g}{2}\sum_j [\Sigma_j^z,[\Sigma_j^z,\ov{\rho \otimes \rho}]]
\end{equation}
where we have adopted analogous notation to that of \eqref{eq:NC-rhorho-Lindblad}, i.e., $\Sigma_j^z = Z_j \otimes \openone + \openone \otimes Z_j$. In the strong-noise limit, the dynamics is projected onto the $6^N$ slow subspace $\cS$ spanned by $\{\ket{z_1}\bra{z_1} \otimes \ket{z_2}\bra{z_2},\ket{z_3}\bra{-z_3} \otimes \ket{-z_3}\bra{z_3}: z_i \in \{-1,1\} \}^{\otimes N}$. The first nonvanishing term in perturbation theory for the generator of the strong-noise dynamics of the second moment of the density matrix, which we also refer to as an effective Liouvillian, is given by $\cL_{\text{eff}} = -\cP_{\cS}L_H D^{-1} L_H\cP_{\cS}$.  If we consider the action of $DL_H$ on a single site, we have
\begin{widetext}
\begin{equation}
\begin{split}
&DL_H\left(\ket{z_1}\bra{z_{1'}} \otimes \ket{z_2}\bra{z_{2'}}\right) = \frac{ig}{2} \Bigg([H,\ket{z_1}\bra{z_{1'}}]\otimes \ket{z_2}\bra{z_{2'}} \\
&\times \left((z_2^j-z_{2'}^j)^2+4\left(1-\frac{(z_1^j-z_{1'}^j)^2}{4}\right) -4z_1^j\left(1-\frac{(z_1^j-z_{1'}^j)^2}{4}\right)(z_2^j-z_{2'}^j) \right) \\
&+ \ket{z_1}\bra{z_{1'}} \otimes [H,\ket{z_2}\bra{z_{2'}}] \\
&\times \left((z_1^j-z_{1'}^j)^2+4\left(1-\frac{(z_2^j-z_{2'}^j)^2}{4}\right) -4z_2^j\left(1-\frac{(z_2^j-z_{2'}^j)^2}{4}\right)(z_1^j-z_{1'}^j) \right) \Bigg),
\end{split}
\end{equation}
\end{widetext}
from which it follows that $L_H\cP_{\cS}\left(\ov{\rho \otimes \rho}\right)$ is an eigenstate of $D$ with eigenvalue $4$. The effective Liouvillian can thus be written
\begin{equation} \label{eq:L_eff}
\cL_{\text{eff}} = -\frac{1}{2g} \cP_{\cS} \, \text{ad}_{\cH}^2,
\end{equation}
where we have used the adjoint action notation $\text{ad}_{\cH}(\cdots) := [\cH,\cdots]$.

\begin{figure}
\includegraphics[width=\columnwidth]{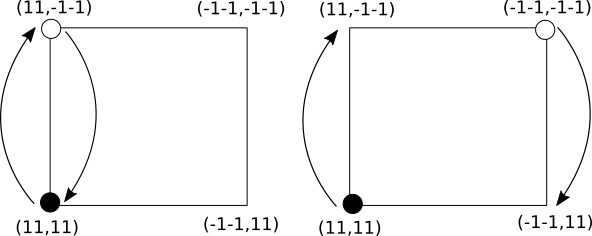}
  \caption{An example of exchange (left) and pair changing (right) terms, as visualised on a square cross section of the `octahedron' of single-site states that span the slow subspace of model C. The filled and empty circles represent the states on adjacent sites.}
  \label{fig:model-c-dynamics}
\end{figure}

In the strong-noise limit, the dynamics is projected onto the $6^N$-dimensional slow subspace $\cS$ spanned by $\{\ket{z_1}\bra{z_1} \otimes \ket{z_2}\bra{z_2},\ket{z_3}\bra{-z_3} \otimes \ket{-z_3}\bra{z_3}: z_i \in \{-1,1\} \}^{\otimes N}$. It is helpful to partition the single-site factors of $\cS$ into three types of pairs of states
\begin{align}
&1. \quad (11,11) \quad &&(-1-1,-1-1) \notag \\
&2. \quad (11,-1-1) \quad && (-1-1,11) \notag \\
&3. \quad (1-1,-11) \quad && (-11,1-1),
\end{align}
where we have represented the state $\ket{z_1}\bra{z_2}\otimes \ket{z_3}\bra{z_4}$ by the tuple $(z_1z_2,z_3z_4)$. It is helpful to visualise each pair of states as occupying antipodal vertices of an octahedron. If we consider the effective Liouvillian, which differs from that of model C only by a multiplicative constant and the fact that the projector is into a different slow subspace, we identify three classes of matrix element (with the possible values given in parenthesis, and their representation on the octahedron given in Fig.~\ref{fig:model-c-dynamics}):
\begin{enumerate}
\item Pair changing ($\pm 2$): A pair of a given type, with each element of the pair occupying adjacent sites, may be transformed into a pair of another type.
\item Exchange ($\pm 2$): The states of adjacent sites may be exchanged, if the two states belong to different pairs.
\item Diagonal ($2$ or $4$): No change, but a constant factor equal to the Hamming distance between the two states is acquired.
\end{enumerate}
The differing signs can be seen to ensure that the both the trace (only pairs 1 and 2 contribute) and purity (only pairs 1 and 3 contribute) of the full density matrix are preserved under evolution.

\section{Effective Liouvillian for the second moment of model NC} \label{sec:Leff-modelNC}
We begin by evaluating the double commutators that arise when $\cL_{\text{eff}}$ acts on the basis states of $\cS$. If we write the Heisenberg Hamiltonian as $\sum_{j,k} h_{jk}$ for $h_{jk} = \sigma_j^a \sigma_k^a$ (n.b. we shall assume the summation convention only for the upper (i.e. spin) indices), we have
\begin{align}
  \left[h_{jk},\left[h_{jk},\sigma_k^a\right]\right]&=8\left(\sigma_k^a-\sigma_j^a\right) \\
\left[h_{jk},\left[h_{jk},\sigma_j^a\sigma_k^b\right]\right]&=8\left(\sigma_j^a\sigma_k^b-\sigma_j^b\sigma_k^a\right).
\end{align}
Since we are considering a Hamiltonian with only a two-body interaction, we only need to consider four possible states: each local factor for the two sites that $h_{jk}$ couples is either $\ket{0}$ or $\ket{1}$ (as defined in \eqref{eq:NC-singlets}). The $\ket{0_j0_k}$ state is trivially seen to lie in the kernel of $\cL_{\text{eff}}$, so that we only need to compute
\begin{align}
  &\cL_{jk} \left(\sigma^a_j\sigma^b_k\otimes \sigma^a_j\sigma^b_k\right) = \cP_{\cS} \Big[-8\left(\sigma_j^a\sigma_k^b-\sigma_j^b\sigma_k^a\right)\otimes \sigma^a_j\sigma^b_k \notag\\
&-8\sigma^a_j\sigma^b_k\otimes\left(\sigma_j^a\sigma_k^b-\sigma_j^b\sigma_k^a\right) +16\left(\sigma^\alpha_k-\sigma^\alpha_j\right)\otimes \left(\sigma^\alpha_k-\sigma^\alpha_j\right) \Big].
\end{align}
and
\begin{align}
&\cL_{jk}\left(\openone_j\sigma^a_k\otimes \openone_j\sigma^a_k\right) = \cP_{\cS} \Big[ -8\left(\openone_j\sigma_k^a-\sigma_j^a\openone_k\right)\otimes\openone_j\sigma^a_k -\notag \\
&8\openone_j\sigma^a_k\otimes\left(\openone_j\sigma_k^a-\sigma_j^a\openone_k\right)
+8\left(\sigma_j^b\sigma_k^c\otimes \sigma_j^b\sigma_k^c-\sigma_j^b\sigma_k^c\otimes \sigma_j^c\sigma_k^b\right) \Big],
\end{align}
with the corresponding result for $\ket{1_j0_k}$ following by interchanging $j$ and $k$ in the last equality. It remains only to perform the projection back into the slow subspace: terms of the form $\sigma_j^a\openone_k \otimes \openone_j\sigma_k^a$ are projected out, but $\sigma_j^b\sigma_k^a\otimes \sigma^a_j\sigma^b_k$ terms have a nonzero component in $\cS$ that we must compute. This is most clearly seen by decomposing the dyadic Cartesian tensor operator $\sigma^a_j \otimes \sigma^b_j$ into irreducible representations of $SO(3)$ as
\begin{align}
 &\sigma^a_j \otimes\sigma^b_j =\frac{1}{3} \left(\sigma^\alpha_j\otimes\sigma^\alpha_j\right)\delta_{ab}+\frac{1}{2}\Big(\sigma^a_j\otimes\sigma^b_j - \sigma^b_j\otimes\sigma^a_j\Big) \notag \\
&+ \frac{1}{2}\left(\sigma^a_j\otimes\sigma^b_j + \sigma^b_j\otimes\sigma^a_j - \frac{2}{3}\sigma^\alpha_j\otimes\sigma^\alpha_j\delta_{ab}\right),
\end{align}
from which it follows that
\begin{equation}
\cP_{\cS}\left[\sigma_j^b\sigma_k^a\otimes \sigma^a_j\sigma^b_k\right] = \frac{1}{3}\sigma_j^a\sigma_k^b\otimes \sigma^a_j\sigma^b_k.
\end{equation}
Combining these results and exploiting orthogonality of the $\{\ket{0},\ket{1}\}$ states with respect to the Hilbert-Schmidt inner product yields the matrix given in \eqref{eq:FA-generator}.

\bibliography{../literature/bibliography.bib}

\end{document}